\documentclass{PoS}

\title{Green's Functions and Topological Configurations}

\ShortTitle{Green's Functions and Topological Configurations}

\author{\speaker{Axel Maas}\thanks{Supported by the DFG under grant number MA 3935/1-2 and by the FWF under grant number P20330. I am grateful to the organizers for inviting me to give this talk, and for this interesting conference.}\\
        Department of Theoretical Physics, Institute of Physics,\\
        Karl-Franzens University Graz, Universit\"atsplatz 5, A-8010 Graz,\\
        Austria\\
        and\\
        Department of Complex Physical Systems, Institute of Physics, \\
        Slovak Academy of Sciences, D\'{u}bravsk\'{a} cesta 9, SK-845 11 Bratislava,\\
        Slovakia\\
        E-mail: \email{Axel.Maas@Uni-Graz.at}}

\abstract{There are, among others, currently two important views on the non-perturbative structure of Yang-Mills theory. One is through topological configurations and one is through Green's functions, in particular their (asymptotic) infrared behavior. Based on both views, various scenarios for confinement, chiral symmetry breaking and other non-perturbative effects have been developed. However, if both views are correct then they can only be different aspects of the same underlying physics, and it must be possible to relate them.

After discussing the current status of the understanding of this connection, smeared and cooled configurations in lattice gauge theory are used to determine the properties of Green's functions in the low-momentum regime. It is found that the qualitative properties are essentially unchanged compared to results on unsmeared configurations. This is also the case when the configurations are smeared sufficiently strongly to reach the almost (anti-)self-dual domain.}

\FullConference{8th Conference Quark Confinement and the Hadron Spectrum \\
		 September 1-6, 2008\\
		 Mainz, Germany}

\begin{document}

\section{Connecting confinement scenarios}

There are two views on the long-distance dynamics of Yang-Mills theory (and QCD), which are currently employed prominently \cite{Alkofer:2006fu}. One view is based on topological configurations of various types \cite{Greensite:2003bk}. Most investigated among them are vortices and monopoles, but there are also other possibilities like merons, calorons, instantons, and others. These configurations provide direct access to the generation of (quark) confinement, chiral symmetry breaking, the finite temperature phase transition and other non-perturbative features of Yang-Mills theory and QCD. The other view is by means of Green's functions \cite{Fischer:2006ub}. These provide access to (gluon) confinement, chiral symmetry breaking, bound states and also other non-perturbative phenomena. They also permit directly the connection to perturbation theory.

Both sets of scenarios are not yet complete, and the discussion on various aspects of them continues \cite{Greensite:2003bk,Fischer:2008uz,Cucchieri:2008yp}. Nonetheless, the most developed scenarios of both cases, the combined vortex and monopole scenario \cite{Alkofer:2006fu,Greensite:2003bk} and the scenarios of Gribov and Zwanziger and of Kugo and Ojima (GZKO) \cite{Alkofer:2006fu,Fischer:2006ub,Fischer:2008uz}, provide already a quite encompassing view of the long-distance shape of QCD.

Given the considerable successes of both views, it seems rather unlikely that either is fundamentally flawed. If therefore both are correct views of the same underlying physics, they must be connected. Understanding this connection may also provide insight into the yet missing pieces of both views, and may also help to settle persisting problems of the individual views.

This relation is to a large extent unexplored. However, there are a number of observations, mostly on the connection of the infrared properties of Landau and Coulomb gauge propagators to vortices, monopoles, and instantons. In this context it is important that most topological configurations are gauge-dependent quantities, and that their identification in various gauges is differing in difficulty \cite{Faber:2001zs}. On the other hand, the GZKO scenario can be applied to the class of linear-$\lambda$ gauges, including the Landau and Coulomb gauge. This requirement of gauge-fixing is due to the concept of quarks and gluons as the basic building blocks of matter, as these are gauge-dependent.

However, at least vortices, instantons \cite{Maas:2005qt}, and monopoles \cite{Maas:2006ss} can be transferred to Landau gauge, although they are not in one-to-one correspondence to their counterparts in other gauges \cite{Greensite:2004ke}. Hence, in the following only Landau (and to some extent Coulomb) gauge will be discussed, despite its complexities \cite{Fischer:2008uz,Maas:2008ri,Sternbeck:2008wg}.

In this gauge it was possible to show that vortices appear inside and on the boundary of the so-called first Gribov region \cite{Maas:2005qt}, monopoles \cite{Maas:2006ss} and instantons \cite{Maas:2005qt} on the boundary of the first Gribov region, and instantons also on the common boundary of the first Gribov region and the so-called fundamental modular region \cite{Maas:2005qt}. According to the Gribov-Zwanziger scenario \cite{Zwanziger:2003cf}, these boundaries are responsible for the infrared properties of Green's functions. Hence, these topological configurations should be relevant field configurations for the Green's functions.

An indirect indication for this is that the spectrum and eigenstates of the Faddeev-Popov operator, which determines the ghost propagator, are affected by topological configurations \cite{Maas:2005qt,Maas:2006ss,Greensite:2004ur}. In particular, they give rise to additional zero-modes, which in principle enhances the ghost propagator. This has also been demonstrated explicitly by determining the ghost \cite{Gattnar:2004bf} and gluon propagators \cite{Langfeld:2002dd} in lattice gauge theory after removal of vortices. In this case, the low-momentum properties are qualitatively altered. However, one of the most surprising results in this context is that instantons also play a role \cite{Maas:2005qt,Boucaud:2003xi}.

\section{Propagators in topological backgrounds}

To investigate this fact more closely, smeared and/or cooled lattice gauge theory configurations can be used. In particular, when smearing sufficiently long, the configurations will be reduced to essentially (anti-)self-dual ones, with an action being almost a multiple of the instanton action.

\begin{figure}
\includegraphics[width=\linewidth]{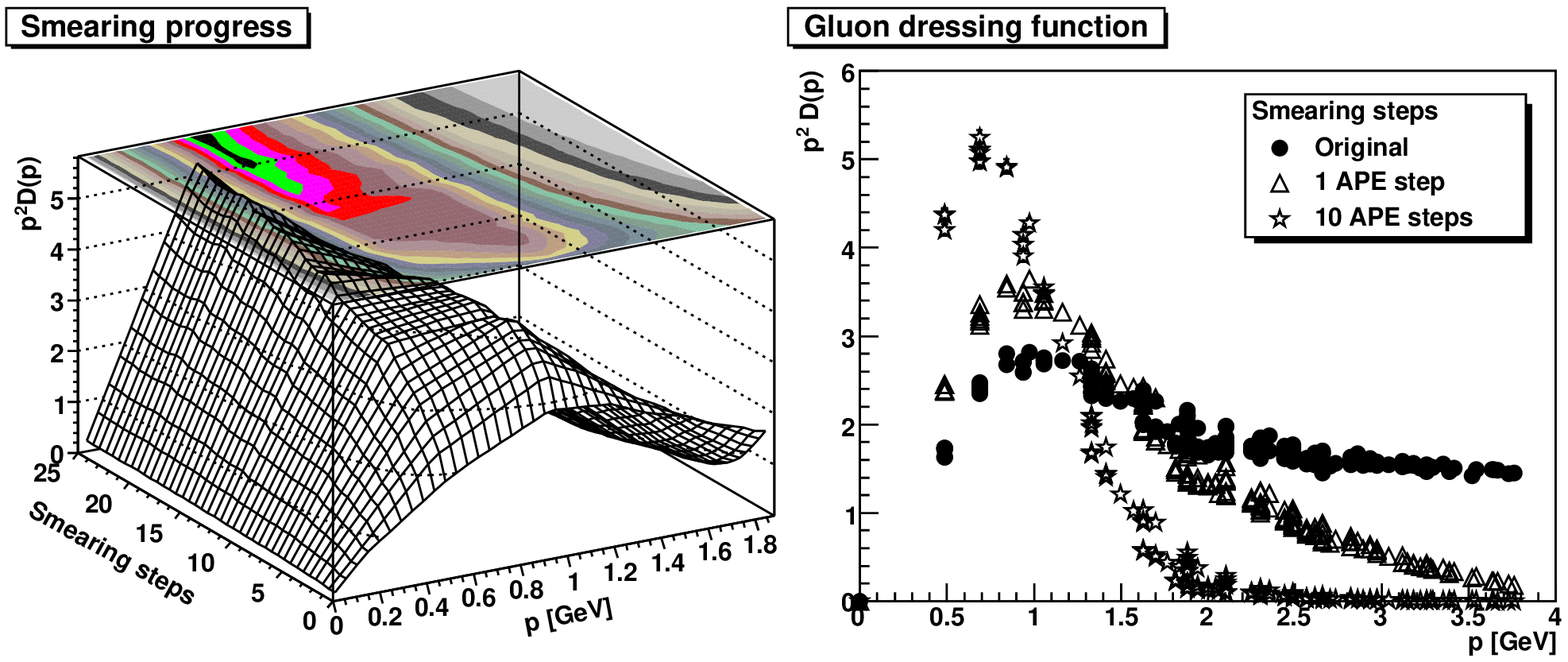}
\includegraphics[width=\linewidth]{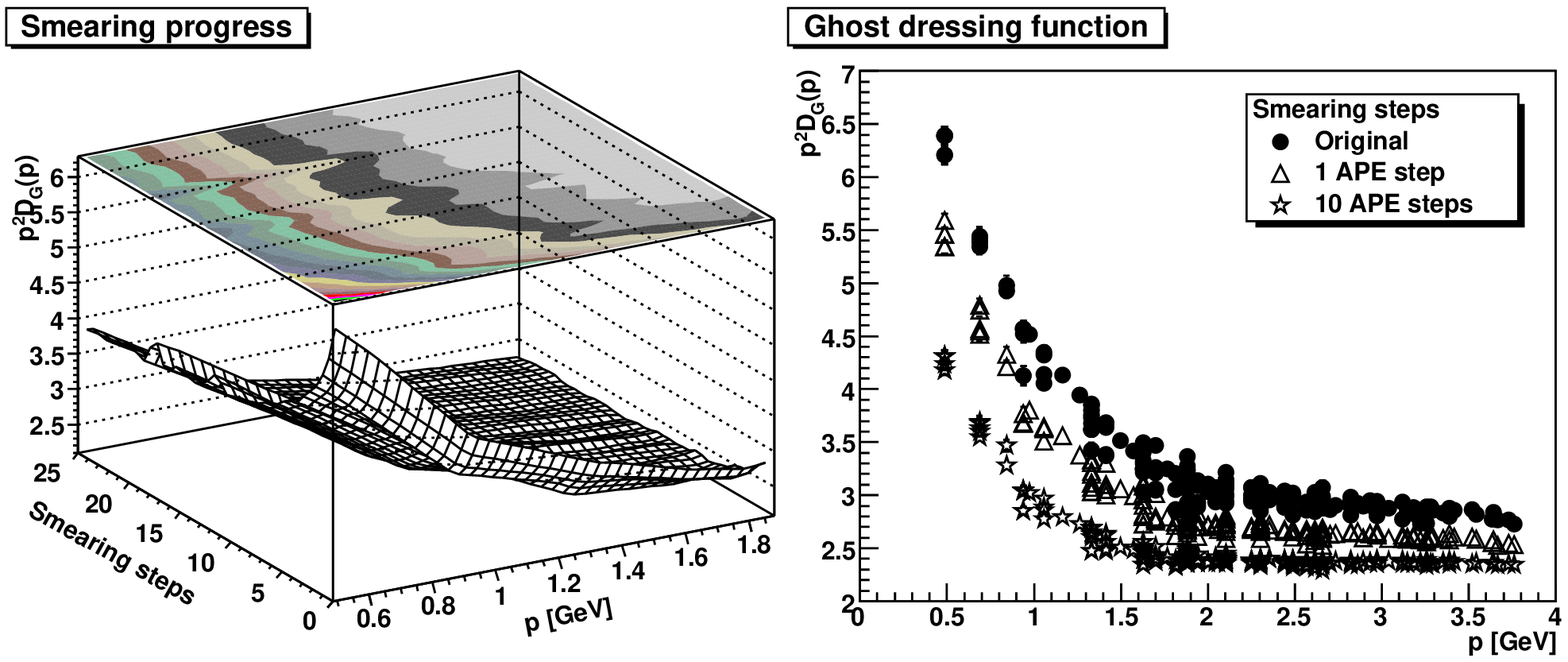}
\caption{The gluon (top) and ghost (bottom) propagators as a function of the smearing step (left panel) after mild smearing and measured after each smearing step. Some individual steps are shown in the right panel. Results are from a $12^4$ lattice at $\beta=2.2$ ($a=0.21$ fm, $V=(2.5$ fm $)^4$)\label{fig1}.}
\end{figure}

\begin{figure}
\includegraphics[width=\linewidth]{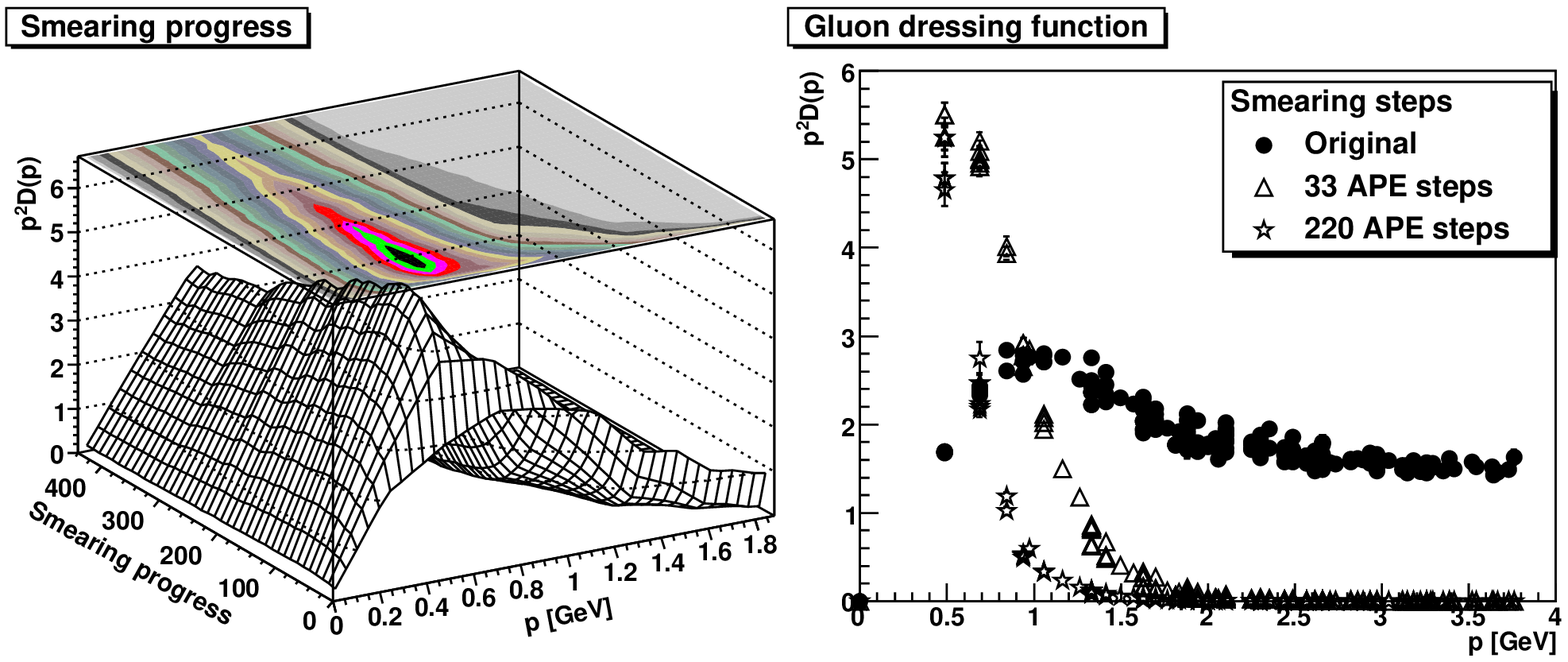}
\includegraphics[width=\linewidth]{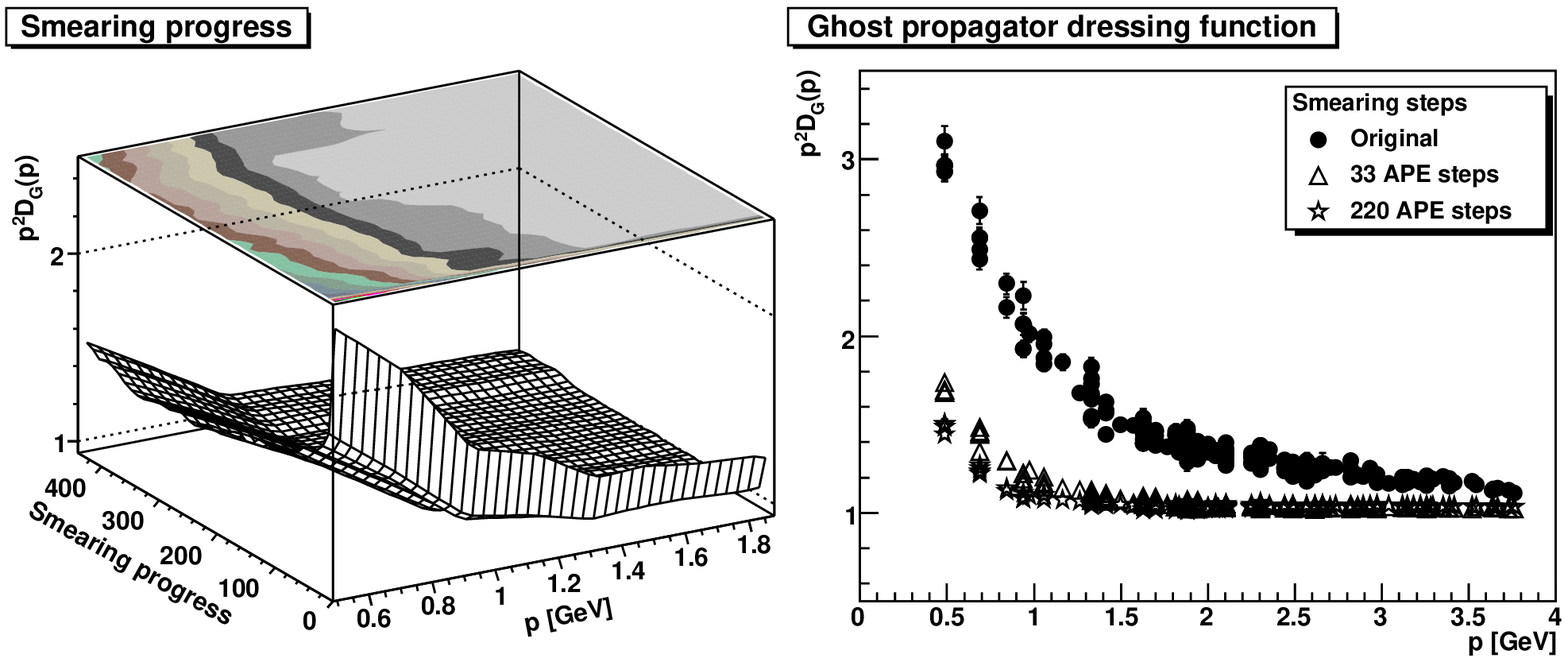}
\caption{Same as in figure 1, but smearing is now extended to 440 APE steps, and measurements are performed every 11 steps. This is deep in the self-dual regime.\label{fig2}.}
\end{figure}

To investigate this in detail, SU(2) lattice configurations in four dimensions \cite{Cucchieri:2006tf} were smeared using a standard APE-smearing \cite{Bruckmann:2006wf}. This procedure is essentially as effective as other methods of smearing for the identification of long-range (topological) structures \cite{Bruckmann:2006wf}. During the smearing, the smeared configurations were fixed to minimal Landau gauge, and the gluon and ghost propagators were determined using standard methods \cite{Cucchieri:2006tf}. By this it was possible to monitor the propagators during the smearing process, which has been done deep into the region of purely topological configurations. Results for rather small lattices, and thus only useful for investigations of the low-momentum properties, are shown in figure \ref{fig1} for mild smearing. In figure \ref{fig2}, results deep in the self-dual regime are shown. A more extended study, for significantly larger volumes and different discretizations, will be shown elsewhere \cite{top}.

The propagators lose their ultraviolet properties under smearing, as expected. The gluon propagator is damped out completely, while the ghost propagator becomes essentially tree-level-like. Their low-momentum properties are, however, not qualitatively affected. Nonetheless, the quantitative effects are considerable, and there might be even stronger effects observable once larger volumes, and thus smaller momenta, can be reached. In fact, the further smearing proceeds, the lower the momentum becomes from which on non-trivial properties of the propagators are observed. The residual configurations, i.\ e., the configurations which the smeared configurations have to be combined with on the group level to obtain the original ones, yield essentially trivial Green's functions, as expected \cite{top}.

It is also possible to determine the propagators in sectors with fixed topological charge. Although this requires large amounts of statistics, preliminary findings indicate that there is no pronounced dependence of the propagators on the topological charge. This will be discussed in detail elsewhere \cite{top}. In principle, such an analysis could also be performed for the instanton number.

\section{Concluding remarks}

The fact that the low-momentum properties of the propagators keep their qualitative properties under smearing provides two insights. On the one hand, this implies that the long-range structure is shaping the low-momentum properties of propagators. This has already been anticipated based on the findings in center-trivial groups \cite{Maas:2007af,Maas:2005ym}. On the other hand, reproducing the low-momentum properties of the Green's functions in functional or other continuum calculations indicates that at least part of the dynamics due to topological configurations has been captured.

Note that these findings are preliminary, and also do not reach far enough into the infrared to make (yet) a statement on the asymptotic properties of the propagators in smeared (topological) field configurations. Nonetheless, the fact that smearing does not qualitatively modify the low-momentum properties of Green's functions is in agreement with the expectations \cite{Maas:2005qt}.


\begin{thebibliography}{99}

\bibitem{Alkofer:2006fu}
  R.~Alkofer and J.~Greensite,
  J.\ Phys.\ G {\bf 34} (2007) S3
  [arXiv:hep-ph/0610365].

\bibitem{Greensite:2003bk}
  J.~Greensite,
  Prog.\ Part.\ Nucl.\ Phys.\  {\bf 51} (2003) 1
  [arXiv:hep-lat/0301023].

\bibitem{Fischer:2006ub}
  C.~S.~Fischer,
  J.\ Phys.\ G {\bf 32} (2006) R253
  [arXiv:hep-ph/0605173].

\bibitem{Fischer:2008uz}
  C.~S.~Fischer, A.~Maas and J.~M.~Pawlowski,
  arXiv:0810.1987 [hep-ph].

\bibitem{Cucchieri:2008yp}
  A.~Cucchieri and T.~Mendes,
  arXiv:0809.2777 [hep-lat].

\bibitem{Faber:2001zs}
  M.~Faber, J.~Greensite and {\v{S}}.~Olejn\'ik,
  JHEP {\bf 0111} (2001) 053
  [arXiv:hep-lat/0106017].

\bibitem{Maas:2005qt}
  A.~Maas,
  Eur.\ Phys.\ J.\  C {\bf 48} (2006) 179
  [arXiv:hep-th/0511307].

\bibitem{Maas:2006ss}
  A.~Maas,
  Nucl.\ Phys.\  A {\bf 790} (2007) 566
  [arXiv:hep-th/0610011].

\bibitem{Greensite:2004ke}
J.~Greensite, {\v{S}}.~Olejn\'ik and D.~Zwanziger,
Phys.\ Rev.\ D {\bf 69} (2004) 074506
[arXiv:hep-lat/0401003].

\bibitem{Maas:2008ri}
  A.~Maas,
  arXiv:0808.3047 [hep-lat].

\bibitem{Sternbeck:2008wg}
  A.~Sternbeck and L.~von Smekal,
  PoS {\bf LATTICE2008} (2008) 267
  [arXiv:0810.3765 [hep-lat]].

\bibitem{Zwanziger:2003cf}
D.~Zwanziger,
Phys.\ Rev.\ D {\bf 69} (2004) 016002
[arXiv:hep-ph/0303028] and references therein.

\bibitem{Greensite:2004ur}
J.~Greensite, {\v{S}}.~Olejn{\'{i}}k and D.~Zwanziger,
JHEP {\bf 0505} (2005) 070
[arXiv:hep-lat/0407032].

\bibitem{Gattnar:2004bf}
J.~Gattnar, K.~Langfeld and H.~Reinhardt,
Phys.\ Rev.\ Lett.\  {\bf 93} (2004) 061601
[arXiv:hep-lat/0403011].

\bibitem{Langfeld:2002dd}
K.~Langfeld, et al.\
arXiv:hep-th/0209173;
J.~Gattnar, et al.\
Nucl.\ Phys.\ B {\bf 716} (2005) 105
[arXiv:hep-lat/0412032].

\bibitem{Boucaud:2003xi}
  P.~Boucaud, F.~De Soto, A.~Le Yaouanc, J.~P.~Leroy, J.~Micheli, O.~Pene and J.~Rodriguez-Quintero,
  Phys.\ Rev.\  D {\bf 70}, 114503 (2004)
  [arXiv:hep-ph/0312332].

\bibitem{Cucchieri:2006tf}
  A.~Cucchieri, A.~Maas and T.~Mendes,
  Phys.\ Rev.\  D {\bf 74}, 014503 (2006)
  [arXiv:hep-lat/0605011].

\bibitem{Bruckmann:2006wf}
  F.~Bruckmann, et al.,
  Eur.\ Phys.\ J.\  A {\bf 33} (2007) 333
  [arXiv:hep-lat/0612024].

\bibitem{top}
  A.~Maas, in preparation.

\bibitem{Maas:2007af}
  A.~Maas and {\v{S}}.~Olejn\'ik,
  JHEP {\bf 0802} (2008) 070
  [arXiv:0711.1451 [hep-lat]].

\bibitem{Maas:2005ym}
  A.~Maas,
  Mod.\ Phys.\ Lett.\  A {\bf 20} (2005) 1797
  [arXiv:hep-ph/0506066].

\end{thebibliography}
\end{document}